# Real Time On Sensor Gait Phase Detection with 0.5KB Deep Learning Model

Yi-An Chen[1], Jien-De Sui[2] and Tian-Sheuan Chang[3], *Member, IEEE*
Instituter of Electronics, National Chiao Tung University, Hsinchu, Taiwan
Contact emails:[1] yianchen.ee06@nctu.edu.tw, [2] vigia117.ee06g@nctu.edu.tw, [3] tschang@g2.nctu.edu.tw

*Abstract*--Gait phase detection with convolution neural network provides accurate classification but demands high computational cost, which inhibits real time low power on-sensor processing. This paper presents a segmentation based gait phase detection with a width and depth downscaled U-Net like model that only needs 0.5KB model size and 67K operations per second with 95.9% accuracy to be easily fitted into resource limited on sensor microcontroller.

## I. Introduction

Gait phase detection on stance and swing time is an essential parameter for personal health care or sports training to derive other gait parameters. Among all methods, Inertial-Measurement-Unit (IMU) sensor based gait phase detection provides a flexible and low cost solution compared to other optical or force plates based methods. Conventional IMU based methods assume zero velocity and accelerations in the stance time that uses threshold or hidden Markov Model for gait event detection. However, these algorithms need per sensor or per person calibration and thus are not general enough for different situations. Recently, with the popularity of deep learning, convolutional neural network (CNN) or recurrent neural network (RNN) have been applied to gait phase detection due to their superior performance. They split the sensor data into segments and classify each segment into one phase, which needs a careful sliding window setting or input delay adjustment to fit different personal speed. Another drawback of the deep learning model is its large model size and corresponding computational complexity that needs high computing power edge devices.

In this paper, we propose a segmentation based gait phase detection that enables per sample point gait phase detection for better timing accuracy, and avoids personal calibrations due to fine grained classification. The presented design is based on our previous proposed model that is a U-Net like approach [3] that has reduced the model parameter numbers size from the original 29M size in the U-Net to 487K. However, such size is still too large to be fitted to the widely used microcontrollers. Flash memory size could be as small as 32KB, which provides very limited resource for model storage beyond the program. In addition, on-chip SRAM is as small as 8KB and up to 320KB, which limits the available run time intermediate feature map size. If the model is too large to be deployed on the sensor microcontroller, all these sensor data has to be transmitted to a powerful edge device such as mobile phone for real time computation, which consumes extra transmission power. In this paper, we adopt width and depth scaling to prune the network first and then train it so that the network size can be reduced to the desired 0.5 KB and still achieves 95.90% accuracy with 8.51 ms and 9.75 ms errors for the swing time and stance time. Such small size of the network allows us to compute sensor data on chip immediately and only need to transmit the result to other device to visualize result, which is both efficient and power-saving.

## II. Methodology

### A. Sensor and Data Collection

This paper collect data from the sensor as shown in Fig. 1 [2], containing a Bluetooth chip, Nordic nRF52832, for data transmission and 9-axis IMU, MPU-9250. In which, only the Bluetooth chip consists of a 64MHz ARM Cortex-M4F with 512KB on-chip flash and 64KB RAM. We intend to run our model on this Bluetooth chip with minimum computation and SRAM requirement to save precious power of the IMU system.

Three normal subjects with ages from 20 to 30 years old are chosen for the experiments. The test tasks includes running and walking with various speeds (5, 7, 9, 11, 13, 17 and 19 km/h). The ground truth data is collected by an optical measurement system. The amount of the whole raw dataset is 1593000 sample-points (1593 seconds with 1000 Hz sampling rate), and downsampled to 50Hz for lower data processing rate.

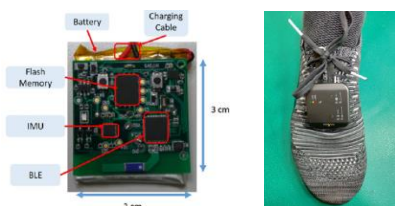

Fig. 1. Position of sensor.

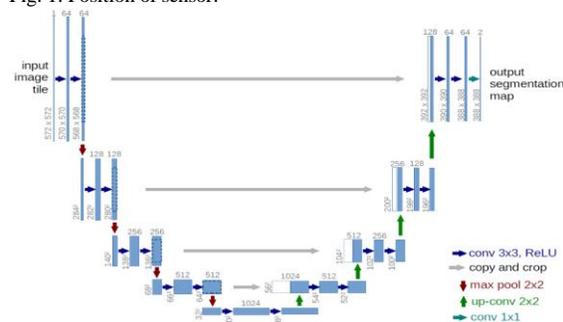

Fig 2. Structure of U-Net, which consists of five layers of the encoder and five layers of the decoder.

## B. Model and Compression with Width and Depth Scaling

The proposed model is based on U-Net as shown in Fig. 2 with only 6-axis sensor data (tri-axis accelerometer and tri-axis gyroscope) as input. This model needs 29M parameter size, which is too large for our Bluetooth chip to store and execute.

The original U-Net model is for image segmentation, which is overparameterized for sensor data processing. To reduce model size, one common way is to use model pruning that prune the unimportant weight or filters and retrain the pruned filter to restore accuracy loss. However, such approaches have to retrain the model and remove the pruned one to get the final clean model, which is time consuming. Besides, in most of these pruning approaches, network depth will be kept unchanged to avoid model collapse.

To avoid above limitations, we adopt the width and depth scaling on the U-Net model directly. This is inspired by the work [4] that training the target network size from scratch with longer training time can have similar performance as complicated pruning procedure. The width scaling is to shrink the model width uniformly. The depth scaling is to shrink the model depth. After model scaling, we further quantize the model to 8-bits so as to use integer arithmetic units in the microcontroller.

## III. EXPERIMENT RESULT

### A. Model Training

The model uses cross entropy as the loss function with mini-batch training (batch size 100) for 500 epochs. This model uses the Adam optimizer with default settings, and sets the learning rate to 0.01. For validation purposes, 3-fold cross validation with independent subjects on the training dataset and testing dataset is applied. 10% of training dataset is used for validation. The model accuracy is per sample point comparison with ground truth as in the segmentation work.

### B. Result

TABLE I
Performance of model with different filter size (comparing to the original model) and different number of layer.

| Number of Layer | Filter Size | Number of Param. | Size(KB) | FLOPs (k/sec) | Accuracy |
|---|---|---|---|---|---|
| 5 | 100% | 29.6 M | 115,814.4 | 1.04M | 95.26% |
| 5 | 12.5% | 29,542 | 112.64 | 1.06k | 96.34% |
| 5 | 6.25% | 7,591 | 30.72 | 291.52 | 94.83% |
| 4 | 6.25% | 1,999 | 10.24 | 154.14 | 95.89% |
| 3 | 6.25% | 510 | 2 | 67.68 | 95.90% |
| 3(quantized) | 6.25% | 510 | 0.5 | 67.68 | 95.50% |

Table 1 shows the result of the model width scaling. The result shows that the accuracy drop is only 0.43% even when downscaling width to 6.25% of the original U-Net. The scaled model only needs 7.5K parameter size (~30.72KB) and 291KFLOPS. Although this seems to be feasible for the target Bluetooth chip, this lefts small margins for other Bluetooth programs. Table 1 shows that the model with 3 layers of depth can remain 95.89% of accuracy, which is slightly higher than the original one due to less overfitting effect. The parameter size has decreased to 510 (2 KB) before quantization. Further quantized the model from 32-bit floating point to 8-bit integer can further reduce the model size to 0.5KB with similar accuracy. The required operation count is 67K. Both these two requirements can be easily meet in current microcontroller, even low end ARM Cortex M0.

### C. Comparison with Other Approaches

Table 2 shows the comparison with other approaches. Since these approaches are not for per sample accuracy, we use commonly used stance and swing time error (average timing error per step for stance and swing phase) as index. The per sample segmentation approach can achieve lower error (at least 50% of error reduction) compared to others.

TABLE II
Comparison of previous studies and our method.

| Author | Stance Time Error | Swing Time Error |
|---|---|---|
| Fabio A. Storm [5] | 30±10ms | None |
| J.-Y. Jung [6] | 21.6ms (continuous error) | |
| R. L. Evans [7] | 23ms (mean error) | |
| **Our Method** | 9.75±11.19ms | 8.51±9.42ms |

## IV. CONCLUSION

This paper presents a real time gait phase detection with a U-Net like segmentation model. The model is optimized with proper model width and depth scaling and quantization so that only needs 0.5 KB model size and 67.68K operation count with 95.90% accuracy with just 8.51 ± 9.42 ms in swing error and 9.75 ± 11.19 ms in stance error. This can be easily implemented on microcontrollers in the IMU sensor kits to achieve real time performance without little computation overhead.

## REFERENCES

[1] Y. Zhang, N. Suda, L. Lai and V. Chandra, "Hello Edge: Keyword Spotting on Microcontrollers", arXiv:1711.07128v3, 2018J. Clerk Maxwell, *A Treatise on Electricity and Magnetism,* 3rd ed., vol. 2. Oxford: Clarendon, 1892, pp. 68-73.
[2] Y. S. Liu and K. A. Wen, "High precision trajectory reconstruction for wearable system using consumer- grade IMU," Tech. Report., May. 2018.
[3] J.-D. Sui, W.-H. Chen, T.-Y. Shiang and T.-S. Chang, "Real-Time Wearable Gait Phase Segmentation For Running And Walking", ISCAS, Oct. 2020
[4] Z. Liu, M. Sun, T. Zhou, G. Huang, and T. Darrell, "Rethinking the value of network pruning", ICLR, 2019
[5] F. A. Storm, C. J. Buckley, and C. Mazz, "Gait event detection in laboratory and real life setting: Accuracy of ankle and waist sensor based methods," Gait & posture, vol. 50, pp. 42-46.
[6] J.-Y.Jung, W. Heo, H. Yang, and H. Park, "A neural network-based gait phase classification method using sensors equipped on lower limb exoskeleton rebots," Sensors, vol. 15, pp. 27738-27759.
[7] R. L. Evans and D. K. Arvind, "Detection of gait phases using orient specks for mobile clinical gait analysis," in 2014 11th International Conference on Wearable and Implementable Body Sensor Networks, pp. 149-154.